\documentclass[12pt]{article}
\usepackage{amsxtra,amssymb,amsthm,amsmath,latexsym}

\theoremstyle{plain}
\newtheorem{theorem}{Theorem}
\newtheorem{lemma}{Lemma}
\newtheorem{remark}{Remark}

\newcommand{\refT}[1]{Theorem~\ref{T:#1}}
\newcommand{\refS}[1]{Section~\ref{S:#1}}

\newcommand{\refL}[1]{Lemma~\ref{L:#1}}

\def\bysame{\rule{.5in}{.005in},\ }

\def\ve{{\varepsilon}}
\def\R{{\mathbb R}}

\def\C{{\mathbb C}}

\def\oH{{\overset{\circ}{H}}}
\def\oH1{{\overset{\circ}{H}\kern-.02in{}^1}}
\def\l{\ell}

\def\Im{{\hbox{\,Im\,}}}
\def\Re{{\hbox{\,Re\,}}}

\def\bee{\begin{equation*}}
\def\eee{\end{equation*}}
\def\be{\begin{equation}}
\def\ee{\end{equation}}
\def\d{\delta}

\begin{document}
\title{Inverse scattering problem with data at fixed energy 
and fixed incident 
direction}

\author{A.G. Ramm\\
 Mathematics Department, Kansas State University, \\
 Manhattan, KS 66506-2602, USA\\
ramm@math.ksu.edu,\\ fax 785-532-0546, tel. 785-532-0580}

\date{}
\maketitle\thispagestyle{empty}

\begin{abstract}
\footnote{key words: inverse scattering, properties of scattering 
amplitudes, quantum mechanics }

Let $A_q(\alpha',\alpha,k)$ be the scattering amplitude, corresponding to
a local potential $q(x)$, $x\in\R^3$, $q(x)=0$ for $|x|>a$, where $a>0$ is 
a fixed number, $\alpha',\alpha\in S^2$ are unit vectors, $S^2$ is the 
unit sphere in $\R^3$, $\alpha$ is the direction of the 
incident wave, 
$k^2>0$ is the energy. We prove that given an arbitrary function 
$f(\alpha')\in L^2(S^2)$,  an arbitrary fixed 
$\alpha_0\in S^2$, an 
arbitrary fixed $k>0$, and an arbitrary small $\ve>0$, there exists a 
potential $q(x)\in L^2(D)$, where $D\subset R^3$ is a bounded domain such 
that
\bee
 \|A_q(\alpha',\alpha_0,k)-f(\alpha')\|_{L^2(S^2)}<\ve. \tag{$\ast$}\eee
The potential $q$, for which $(\ast)$ holds, is nonunique. We give a 
method for finding $q$, and a formula for such a $q$.

\end{abstract}


\section{Introduction}\label{S:1}
If $q(x)=0$ for $|x|>a$, where $a>0$ 
is some number, $q(x)\in L^2(B_a)$, $B_a=\{x:|x|\leq a\}$, then the 
corresponding scattering amplitude $A_q(\alpha',\alpha,k)$ is defined as 
follows. Let $\alpha\in S^2$ be a given unit vector, where $S^2$ is the 
unit 
sphere. The scattering problem consists in finding 
the scattering solution 
$u(x,\alpha,k)$, which solves the equation
\be\label{e1}
 [\nabla^2+k^2-q(x)]u=0\hbox{\quad in\quad} \R^3, \ee
and satisfies the following asymptotics:
\be\label{e2}
 u=e^{ik\alpha\cdot x} +A_q(\alpha',\alpha,k)\frac{e^{ikr}}{r} 
 +o\left(\frac{1}{r}\right),\quad r:=|x|\to\infty, \quad 
 \alpha':=\frac{x}{r}. \ee
Vector $\alpha$ is called the incident direction, the direction of the 
incident plane wave $e^{ik\alpha\cdot x}$, vector $\alpha'$ is the 
direction in which the incident wave is scattered, and the coefficient 
$A_q(\alpha',\alpha,k)$ is called the scattering amplitude. The properties 
of the scattering amplitude, corresponding to a real-valued, rapidly 
decaying at infinity $q$, has been studied in detail in the literature. 
(See, e.~g.,~\cite{C}).

The inverse scattering problem with fixed-energy data consists in finding
a potential $q(x)$ from the knowledge of $A_q(\alpha',\alpha,k)$ at a
fixed $k>0$ and all $\alpha',\alpha\in S^2$. The uniqueness of the
solution to this probem in the class of real-valued, compactly supported,
square-integrable potentials was first proved by the author (\cite{R228}),
who also gave a method for recovery of $q$ from the exact fixed-energy
data, an error estimate for this method, and stability estimates for the
recovery problem (\cite{R470}) as well as a method for a stable recovery
of $q$ from noisy fixed-energy data, and an error estimate for this method
\cite{R285}, \cite{R425}, \cite{R470}. Until now this is the only
known rigorously justified method for recovery of $q$ from noisy
fixed-energy data.

In this paper a new problem is studied. Let us assume that the incident 
direction $\alpha$ is fixed, $\alpha=\alpha_0$, and $k>0$ is also fixed, 
$k=k_0>0$. Denote by $A_q(\alpha'):=A_q(\alpha',\alpha_0,k_0)$
the corresponding scattering amplitude, and let $\beta:=\alpha'$. Let 
$f(\beta)\in L^2(S^2)$ be an arbitrary function.
In part of our arguments, in Lemma 3, specifically, it is convenient to 
assume first that  
the norm of $f$ is sufficiently small 
norm $||f(\beta)||_{ L^2(S^2)}$. The role of this "smallness" 
assumption will be clear from our 
arguments.  Under this "smallness" assumption we derive an analytical 
explicit formula for the potential which generates the scattering 
amplitude at a fixed $k>0$ and a fixed incident direction $\alpha$
with any desired accuracy. The 
"smallness" assumptions will be dropped in Lemma 4.

Fix an arbitrary small 
number $\ve>0$. Let $D\subset \R^3$ be a bounded domain. 

The problem (P) is: 

{\it Given $\ve>0$, $f(\beta)\in L^2(S^2)$, $||f(\beta)||_{ L^2(S^2)}$, 
$\alpha_0\in S^2$, and 
$k_0>0$, 
does there exist a potential $q\in L^2(D)$ such that 
\be\label{e3} 
 \|f(\beta)-A_q(\beta)\|_{L^2(S^2)}<\ve, \ee
and how does one calculate such a $q$?}

Our basic result consists of an answer to these questions.
In \cite{R512} the approximation problem problem similar to
the above was considered for the first time, but the argument in 
\cite{R512} does require the "smallness" assumption. 

\begin{theorem}\label{T:1}
For any $f\in L^2(S^2)$, an arbitrary 
small $\ve>0$, any fixed 
$\alpha=\alpha_0\in S^2$, any fixed $k=k_0>0$, 
and any bounded domain $D\subset \R^3$,
there exists a (non-unique) potential $q(x)\in L^2(D)$, such that 
\eqref{e3}
holds. \end{theorem}

To calculate such a potential, we need some auxiliary results.

\begin{lemma}\label{L:1}
Given an arbitrary $f(\beta)\in L^2(S^2)$ and an arbitrary small $\ve>0$, 
there exists an $h\in L^2(D)$, such that \end{lemma}
\be\label{e4}
 \bigg\|f(\beta)+\frac{1}{4\pi}\int_D e^{-ik\beta\cdot x} 
h(x)dx\bigg\|<\ve. 
\ee

In Lemma 1, and in Lemma 2 below, we do not need the "smallness" 
assumption. In what follows we 
always assume that $\alpha=\alpha_0\in S^2$ and 
$k=k_0>0$ are fixed, and write $\alpha,k$ in place of $\alpha_0$ and 
$k_0$.

The following formula for the scattering amplitude is well known: 
\be\label{e5}
 A_q(\beta,\alpha,k)=-\frac{1}{4\pi}
 \int_D e^{-ik\beta\cdot x} q(x)u(x)dx,\ee
where $u(x):=u(x,\alpha,k)$ is the scattering solution and the dependence 
on $\alpha$ and $k$ is not shown because $\alpha$ and $k$ are fixed.

If
\be\label{e6}    h(x)=q(x)u(x), \ee
then \eqref{e4} is identical to \eqref{e3}.

We now explain how one calculates $q$ from equation
\eqref{e6} if $h$ is given, and how one calculates $h$,
satisfying \eqref{e4}, if $f$ is given.

\begin{lemma}\label{L:2}
Given $f\in L^2(S^2)$, one calculates $h$, satisfying \eqref{e4}, by the 
formula:
\be\label{e7}
 h_{\l,m}=\left\{\begin{array}{ll}-(-i)^\l 
 \frac{f_{\l,m}}
      {\sqrt{\frac{\pi}{2k}} g_{1,\l+\frac12}(k)}, & \l\leq L,\\
 0\hfill, & \l>L\end{array}\right. \ee
where $g_{\mu,\nu}(k):=\int^1_0 x^{\mu+\frac12}J_\nu(kx)dx$. This integral 
can be calculated analytically {\rm (\cite{B}, formula 8.5.8)}, and we 
have assumed 
that $h(x)=0$ for $|x|>1$.
\end{lemma}

If $h$ is found such that \eqref{e4} holds, then $q$ can be
calculated from equation \eqref{e6}, which is a nonlinear
equation for $q$, because the scattering solution $u$
depends on $q$.  Under the "smallness" assumption
we derive an analytical explicit formula for $q$, namely,
formula \eqref{e9} below. This formula holds 
not necessarily under the "smallness"  
assumption, as follows from Lemma 4.

In Lemma 3 we use the "smallness" assumption for the
first time. This assumption guarantees that inequality
\eqref{e8} holds. This inequality is sufficient for the
formula \eqref{e9} to produce an $L^2(D)$ potential $q$. 

In Lemma 4 we prove that there exists an $h_\d$, a small
perturbation of $h$, $||h-h_\d||_{L^2(D)}<\d$, which leads
by formula \eqref{e9}, with $h$ replaced by $h_\d$, to a
potential $q_\d$, generating the radiation pattern
$A(\beta)$ as close to the desired $f$ as one wishes.

\begin{lemma}\label{L:3}
Assume that $\sup_{x\in D}\bigg| \int_D ghdy\bigg|<1$, or, more generally,
that
\be\label{e8}
 \sup_{x\in D}\bigg|u_0(x)- \int_D g(x,y)h(y)dy\bigg|>0,\qquad 
 g=g(x,y):=\frac{e^{ik|x-y|}}{4\pi|x-y|}. \ee
Then equation \eqref{e6} has a unique solution:
\be\label{e9}
 q(x)=\frac{h(x)}{u_0(x)-\int_D g(x,y)h(y)dy}, \qquad 
 u_0(x)=e^{ik\alpha\cdot x}, \quad q\in L^2(D). \ee
Formula \eqref{e9} holds also if condition \eqref{e8} holds.
\end{lemma}

Suppose that for a given $h\in L^2(D)$ condition \eqref{e8} 
is not satisfied. Let us  approximate $h$ by an analytic 
function $h_1$ in $D$, for example, by a polynomial, so that
  $$||f(\beta)+\frac{1}{4\pi}\int_D e^{-ik\beta\cdot 
x} h_1(x)dx||<\ve.$$ 
Denoting $h_1$ by $h$ again, we may assume that $h$ is 
analytic in $D$ and in a domain which contains $D$.
In the following lemma we prove that it is possible
to perturb $h$ slightly so that for the perturbed $h$,
denoted $h_\d$, condition \eqref{e8}
is satisfied, and formula \eqref{e9} yields a potential 
$q_\d \in L^2(D)$, for  which inequality \eqref{e3}
holds. 

\begin{lemma}\label{L:4} Assume that $h$ is analytic in $D$ 
and bounded in the closure of $D$. There exists a small 
perturbation $h_\d$ of $h$,
$||h-h_\d||_{L^2(D)}<\d$, such that the function
$$q_\d:=\frac{h_\d(x)}{u_0(x)-\int_Dg(x,y)h_\d(y)dy}$$ is 
bounded.
\end{lemma}

In \refS{2}	 proofs are given.

\section{Proofs}\label{S:2}

\begin{proof}[Proof of \refL{1}]
Assume the contrary. Then
\bee \int_{S^2} d\beta f(\beta) \int_D e^{-ik\beta\cdot x} 
  h(x)dx=0 \qquad \forall h\in L^2(D). \eee
This implies
\be\label{e10}
 \int_{S^2} f(\beta)e^{-ik\beta\cdot x} d\beta=0 
\qquad x\in D. \ee

The left-hand side of\eqref{e10} is the Fourier transform of a 
distribution $f(\beta) \frac{\delta(\lambda-k)}{\lambda^2}$, where 
$\delta(\lambda-k)$ is the delta-function and $\lambda\beta$ is the 
Fourier transform variable in spherical coordinates. By the injectivity of 
the Fourier transform, 
\eqref{e10} implies that $f(\beta)=0$.
\end{proof}

\begin{remark}\label{R:1}
In this proof and in the proof of \refL{2} 
below, the domain $D$ can be 
taken such that $diam D$ is arbtrarily small.
\end{remark}

\begin{proof}[Proof of \refL{2}]
Note that \refL{2} implies \refL{1}. It is known that
\be\label{e11}
  e^{-ik\beta\cdot x}=\sum^\infty_{\l=0} 4\pi(-i)^\l j_\l(kr)
  \overline{Y_{\l,m}(\alpha')} Y_{\l,m}(\beta), \ee
where $\sum_\l=\sum_\l\ \sum^\l_{m=-\l}$, 
$j_\l(r)=\sqrt{\frac{\pi}{2r}} J_{\l+\frac12} (r)$, 
$J_\l(r)$ is the Bessel function, $Y_{\l,m}$ are orthonormal in $L^2(S^2)$ 
spherical harmonics, $Y_{\l,m}(-\beta)=(-1)^\l Y_{\l,m}(\beta)$, 
$\alpha'=\frac{x}{r}$, $r=|x|$, and the overbar stands for complex 
conjugate. 
Expand $f$ in the Fourier series: 
\[f(\beta)=\sum^\infty_{\l=0} f_{\l,m} Y_{\l,m}(\beta).\]
Choose $L$, such that
\be\label{e12}  \sum_{\l>L} |f_{\l,m}|^2 <\ve^2. \ee
Let $\l>L$, $h(x)=\sum^\infty_{\l=0}h_{\l,m}(r) Y_{\l,m}(\alpha')$,
$\alpha'=\frac{x}{r}$, $r=|x|$.

Define $h_{\l,m}$ by \eqref{e7}. Then, by Parseval's equation, inequality 
\eqref{e4} holds. \refL{2} is proved.
\end{proof}

\begin{remark}\label{R:2}
Alternatively, one may look for $h$ in the form $h=\sum^J_{j=1} c_j 
\varphi_j(x)$, where $(\varphi_j,\varphi_i)=\delta_{ij}$ and $c_j=const$, 
and minimize the left-hand side of \eqref{e4} with respect to $c_j$, 
$1\leq j\leq J$. If $J$ is sufficiently large, the minimum will be 
$\leq\ve$. One may take not an orthonormal system of 
$\varphi_j$, but just a linearly independent, complete (total) in 
$L^2(D)$, system of functions $\{\varphi_j\}$.
\end{remark}

\begin{proof}[Proof of \refL{3}]
The scattering solution, corresponding to a potential $q$, solves the 
equation
\be\label{e13}
 u=u_0-\int_D g(x,y)q(y)u(y)dy, \qquad u_0:=e^{ik\alpha\cdot x}.\ee
If \eqref{e6} holds, i.e., if $h$ corresponds to a $q\in L^2(D)$, then 
$u=u_0-\int_D ghdy$. Multiply this equation 
by $q$ and get \[q(x)u(x)=q(x)u_0(x)-q(x)\int_D g(x,y)h(y)dy.\] Using 
\eqref{e6} and solving for 
$q$, 
one gets \eqref{e9}, provided that \eqref{e8} holds.
Condition  \eqref{e8} holds if $\|h\|_{L^2(D)}$ is sufficiently small. 
One has
\be\label{e14}
 \bigg|\int_D g(x,y)hdy\bigg|\leq \frac{1}{4\pi} \sup_x 
\bigg\|\frac{1}{|x-y|}\bigg\|_{L^2(D)} \|h\|_{L^2(D)},\ee
and $\frac{1}{4\pi}\left(\int_D \frac{dy}{|x-y|^2}\right)^{\frac12}\leq
 \frac{\sqrt{a}}{\sqrt{4\pi}}$, where $a=\frac12 diam D$. If, for 
example,
$\frac{\sqrt{a}}{\sqrt{4\pi}}\|h\|_{L^2(D)}<1$, then condition \eqref{e8} 
holds, and formula \eqref{e9} yields the corresponding potential.
This explains the role of the "smallness" assumption.

If $h\in L^2(D)$ is an arbitrary function, such that condition \eqref{e8}
holds, then formula \eqref{e9} defines a potential $q\in L^2(D)$,
formula \eqref{e6} defines the function $u=\frac h q$, and this $u$
solves equation \eqref{e13}, which is the equation for the scattering 
solution. Thus, the function 
$u=\frac h q$ is the scattering solution, and the potential $q$,
defined by formula \eqref{e9}, corresponds to the given $h$.
The above argument is valid in the case of complex-valued potential $q$.

By Lemma 4, proved at the end of the paper, if $h$ is such that
condition \eqref{e8} fails, then a small perturbation $h_\d$ of $h$ leads
by formula \eqref{e9}, with $h_\d$ in place of $h$, to a potential
$q_\d$ which generates the radiation pattern close to the desired $f$.

\refL{3} is proved.
\end{proof}

\begin{remark}\label{R:3}
Formula \eqref{e9} yields, possibly, a complex-valued potential. If $k>0$, 
then $k^2>0$ is not an eigenvalue of the Schr\"odinger operator 
$-\nabla^2+q(x)$ with a complex-valued compactly supported $q\in L^2(D)$. 
This follows, e.g., from the results in \cite{K} for 
$q=o\left(\frac{1}{|x|}\right)$ as $|x|\to\infty$,
and can be proved for a compactly supported $q$ by a simple argument. 
\end{remark}

In the proof of \refL{2} one may take $h(x)=\sum^\infty_{\l=0} h_{\l,m} 
Y_{\l,m}\left(\frac{x}{|x|}\right)$, $|x|\leq b<1$, where $h_{\l,m}$ do 
not depend on $r=|x|$, $h(x)=0$ for $|x|>b$, and $b>0$ can be an arbitrary 
small number.

\begin{proof}[Proof of \refT{1}]
Given an arbitrary $f\in L^2(S^2)$ and an arbitrary small $\ve>0$, choose 
$h$, such that \eqref{e4} holds. This is possible by \refL{1} and 
is done analytically in \refL{2}. If such an $h$ is found, then
one calculates $q$ by formula (9), provided that $h$ is 
sufficiently small, i.e., $f(\beta)$ is sufficiently small. If $f(\beta)$ 
is arbitrary, 
so that conditions \eqref{e8} or \eqref{e14} are not satisfied, then
$q$ is found numerically.

Let us explain the role of the "smallness" assumption on $f$.
From \refL{2} it follows that $f$ and $h$ are proportional, so that if 
$||f||_{L^2(S^2)}$ is sufficiently small, then $||h||_{L^2(D)}$ is small, 
and then 
condition  \eqref{e8}  is satisfied. In this case
the potential is given by formula \eqref{e9}. Theorem 1 is proved.
\end{proof}
\begin{remark}\label{R:4}
If \eqref{e8} fail, then formula \eqref{e9} may yield a 
$q\notin L^2(D)$. As long as formula \eqref{e9} yields a potential
$q\in L^p(D),\,\, p\geq 1,$ our arguments essentially remain valid.
In our presentation we have used $p=2$ because the numerical
minimization in $L^2$-norm is simpler.

The difficulty arises when formula \eqref{e9} yields a
potential which is not locally integrable. Numerical
experiments showed that this case did not occur in practice
in several test examples in which the "smallness" condition
was not satisfied. 

We prove that a suitable small perturbation $h_\delta$ 
of $h$ in 
$L^2(D)$-norm yields by formula \eqref{e9} a bounded 
potential $q_\delta$. This means that the "smallness" 
restriction on the norm of $f$ can be dropped.
The proof is given below.

\end{remark}


\begin{proof}[Proof of \refL{4}] Let $$N:=\{x: \psi(x)=0,
x\in D\},$$ where $$\psi:=u_0(x)-\int_Dg(x,y)h_\d(y)dy.$$
This set is generically a line, defined by two simultaneous
equations $\psi_j=0, j=1,2$, where $\psi_1:=\Re \psi$ and
$\psi_2:=\Im \psi$. Let
$$N_\d:=\{x: |\psi(x)|<\d, x\in D\}$$
and $D_\d:=D\setminus N_\d$. Generically, 
$|\nabla \psi|\geq c>0$ on $N$, and, therefore, by
continuity, in $N_\d$. A small perturbation of $h$ will lead
to these generic assumptions. Consider the new coordinates
$$s_1=\psi_1, \quad s_2=\psi_2,  \quad s_3=x_3.$$ 
Choose the origin on
$N$. The Jacobian $\frac {\partial(s_1, 
s_2, s_3)}{\partial(x_1, x_2, x_3)}$ is
non-singular in $N_\d$. The vectors $\nabla \psi_j, \,
j=1,2,$ are linearly independent in $N_\d$. Define $h_\d=h$
in $D_\d$ and $h_\d=0$ in $N_\d$. Let 
$$q_\d:=\frac{h_\d}{\psi_\d},$$ 
where 
$$\psi_\d:=u_0(x)-\int_Dg(x,y)h_\d(y)dy.$$ 
We wish to prove 
that the
function $q_\d$ is bounded. It is sufficient to check that
$|\psi_\d|>c>0$ in $N_\d$. By $c$ we denote various positive
constants independent of $\d$. One has 
$$|\psi_\d|\geq
|\psi|-I(\d)\geq \d-I(\d),$$ where 
$$I(\d):=\frac
M{4\pi}\int_{N_\d}\frac {dy}{|x-y|}, \quad x\in D_\d,\quad
M=\max_{x\in N_\d}|\psi|.$$ 
The proof will be completed if
the estimate $$I(\d)=O(\d^2|\ln(\d)|)$$ 
is established. Let us
derive this estimate. It is sufficient to check this
estimate  for the integral $$I:=\int_{N_\d}\frac
{dy}{|y|}=2\pi\int_{\rho\leq \d}d\rho \rho \int_0^1 \frac
{ds_3}{\sqrt{s_3^2+\rho^2}},$$ where $\rho^2=s_1^2+s_2^2$. A
direct calculation yields the desired estimate:
$I=O(\d^2|\ln(\d)|)$. Lemma is proved. \end{proof}

Finally we make some remarks about ill-posedness of our
algoritm for finding $q$ given $f$. This problem is
ill-posed because an arbitrary $f\in L^2(S^2)$ cannot be the
scattering amplitude $A_q(\beta)$ corresponding to a
compactly supported potential $q$. Indeed, it is proved in
[5], [7], that $A(\beta)$ is infinitely differentiable on
$S^2$ and is a restriction to $S^2$ of a function analytic
on the algebraic variety in $\C^3$, defined by the equation
$\beta\cdot\beta=k^2$. Finding $h$ satisfying (4) is an
ill-posed problem if $\ve$ is small. It is similar to
solving the first-kind Fredholm integral equation
$$\frac{1}{4\pi}\int_D e^{-ik\beta\cdot 
x}h(x)dx=-f(\beta),$$
whose kernel is infinitely smooth. Our solution (7) shows
the ill-posedness of the problem because the denominator in
(7) tends to zero as $\ell$ grows. Methods for stable
solutions of ill-posed problems (see \cite{R470}) should be
applied to finding $h$. If $h$ is found, then $q$ is found
by formula \eqref{e9}, provided that \eqref{e8} holds. If
\eqref{e8} does not hold, one perturbs slightly $h$
according to Lemma 4, and get a potential $q_\d$ by formula
\eqref{e9} with $h_\d$ in place of $h$.


\begin{thebibliography}{1000} 

\bibitem{B}   Bateman, H.,  Erdelyi, A., {\bf Tables of integral
transforms},
McGraw-Hill, New York, 1954.

\bibitem{C} Cycon, H., Froese, R., Kirsch, W., Simon, B.,
{\bf Schr\"odinger operators}, Springer, Berlin, 1986.

\bibitem{K} Kato, T., ``Growth properties of solutions of the reduced 
wave equation with a variable coefficient", Comm.~Pure and Appl.~Math.~, 
12, (1959), 403-425.

\bibitem{R228}
Ramm, A.~G.~,
``Recovery of the potential from fixed energy scattering data". Inverse 
Problems, 4, (1988), 877-886.

\bibitem{R285} 
\bysame
``Stability estimates in inverse scattering'', Acta Appl. Math., 28, N1, 
(1992), 1-42.

\bibitem{R425} \bysame
``Stability of solutions to inverse scattering problems with fixed-energy 
data", Milan Journ of Math., 70, (2002), 97-161.

\bibitem{R470}
\bysame {\bf Inverse problems}, Springer, New York, 2005.

\bibitem{R512} Completeness of the set of scattering amplitudes, Phys. 
Lett. A, (to appear)

\end{thebibliography}
\end{document}